\documentclass[usegraphicx]{mn2e}
\begin{document}

\title[MOND and CDM]{A tensor-vector-scalar framework for modified
dynamics and cosmic dark matter}

\author[R.H. Sanders] {R.H.~Sanders\\Kapteyn Astronomical Institute,
P.O.~Box 800,  9700 AV Groningen, The Netherlands}

 \date{received: ; accepted: }

\maketitle

\begin{abstract}
I describe a tensor-vector-scalar theory that reconciles the galaxy
scale success of modified Newtonian dynamics (MOND) with the cosmological
scale evidence for CDM.  The theory provides a cosmological basis
for MOND in the sense that the predicted phenomenology only arises in
a cosmological background.  The theory contains an evolving effective 
potential, and scalar field oscillations in this potential comprise the
cold dark matter;  the de Broglie wavelength of these soft bosons, however,
is sufficiently large that they cannot accumulate in galaxies.  The theory
predicts, inevitably, a constant anomalous acceleration in the outer
solar system which, depending upon the choice of parameters, can be
consistent with that detected by the Pioneer spacecrafts.

\end{abstract}


\section{Introduction}
The cosmological evidence supporting the existence of a 
universal pressure-less fluid (cold dark matter or CDM) appears to 
be compelling.  The amplitude of the first two acoustic peaks in
the power spectrum of the CMB anisotropies as observed by WMAP implies that
this fluid comprises, at present, about 30\% of the mass density of
the Universe.  Given the even more compelling evidence for
an $\Omega=1$ universe, most of the remainder must consist of a
negative pressure ``dark energy'' component that may be identified with
a positive cosmological constant (Spergel et al. 2003, Page et al. 2003).  
It is now well-known
that type 1a supernovae (Garnevitch et al. 1998, Perlmutter et al. 1999) 
near $z\approx 0.5$ are systematically dimmer by about 
0.2 magnitudes than would be expected in an empty, ``coasting'' Universe.
These observations are often cited as
evidence that the Universe is presently dominated by this vacuum energy.  But, 
if we can trust that systematic effects are well-understood,
it is also true that the relative brightening of SNIa beyond 
$z\approx 1$ (Tonry et al. 2003) is equally strong evidence for 
matter domination at this relatively 
recent epoch, again implying that $\Omega_{CDM} \approx 0.3$.  Several
such independent tests have lead to the emergence of a ``concordance model'' 
model for the Universe (Ostriker and Steinhardt 1995), which is now heralded 
as a major achievement in observational cosmology and in which CDM
is a vital component.

Although the large scale
evidence for CDM is persuasive, the hypothesis is not without problems.
While there is no shortage of candidate particles, 
the nature of this pressure-less fluid remains elusive.  Particle
physics theory beyond the standard model provides no definitive prediction, and
attempts at direct laboratory detection of CDM particles (neutralinos, axions,
etc.) have, so far, proved unsuccessful.  Moreover, the evidence for CDM
on the scale of galaxies is far from compelling, with ``complicated'', but
poorly understood, astrophysical processes being invoked to explain various
observed phenomena such as the tightness of the
luminosity-rotation velocity relation for disk galaxies--
the Tully-Fisher law-- and the presence of an upper limit on the surface
brightness of galaxies-- the Freeman law (Dalcanton et al. 1997,
van den Bosch \& Dalcanton 2000).  The problem of matching the predicted
rotation curves of dark halos emerging from cosmic N-body simulations with
those actually observed in low surface brightness galaxies has provoked
an ongoing controversy (e.g. de Blok et al 2001).

These same phenomena, on a galaxy scale, are well-described by 
an ad hoc modification
of Newtonian dynamics, MOND, suggested by Milgrom (1983). Here it
is proposed that, below a critical acceleration, $a_0$,   
the effective gravitational acceleration, $g$, approaches
$\sqrt{g_n a_0}$ where $g_n$ is the usual Newtonian acceleration.  
This critical acceleration is, within an order of magnitude,
comparable to the velocity of light times the present value of the
Hubble parameter, i.e., $a_0 \approx cH_0/6$.  This simple modification
accounts for systematic
aspects of galaxy photometry and kinematics, such as the Freeman law and the
Tully-Fisher law,
as well as successfully predicting the detailed form of galaxy rotation 
curves from the observed distribution of detectable baryonic matter.
These successes have been well-cataloged (Sanders and McGaugh 2002) and
suggest that galaxy phenomenology can be understood without 
invoking the existence of dark matter (McGaugh 2004a).

MOND is also not without problems.  On a phenomenological level, the
algorithm appears unable to account for the full mass discrepancy
in rich clusters of galaxies;  it remains necessary to invoke the presence
of, as yet, undetected matter in these systems (Sanders 1999, 2003; 
Aguirre et al. 2002).  
Moreover, the idea has long been criticized
as an ad hoc, empirically based hypothesis without foundation in 
deeper theory.  The absence of a viable relativistic theory has placed
the issues of cosmology, structure formation, and gravitational
lensing beyond consideration in the context of MOND;  that is to say, MOND
cannot be confronted by this entire body 
of observed cosmic phenomena.  Recently,
Bekenstein (2004) has written down a fully covariant theory, {\it TeVeS}, 
involving dynamic vector and scalar fields
in addition to the usual tensor field of General Relativity.  This 
is an important development because the theory
is free of the conspicuous pathologies of earlier attempts (e.g., acausal
wave propagation, no enhanced gravitational lensing).  However, in the
theory as it now stands,
there is no obvious connection between MOND phenomenology and
cosmology as is suggested by the near coincidence between $a_0$ and $cH_0$.
The free function of the theory has two discontinuous
branches-- one for cosmology and one for mass concentrations;  it would
seem difficult to follow the cosmic growth of inhomogeneities in the context 
of a theory in which cosmology is mathematically disconnected 
from inhomogeneous structure.
Finally, the theory does not address what is perhaps the greatest
problem for MOND:  the large scale evidence for CDM cited above.

The construction of a cosmologically effective theory of MOND, and,
in particular,
the reconciliation of the cosmological evidence for CDM with the 
galaxy-scale success of MOND, is the topic of the present paper.  
I will show here that this
is a possibility in the context of a generalization of an earlier
theory also due to Bekenstein (1988a)-- phase coupling gravity or PCG--
but a generalization that requires a vector field as in TeVeS.

The basic idea is this:  PCG is a scalar-tensor theory with two
scalar fields, one of which couples to matter, the second determining
the strength of that coupling.  In a cosmological context, the scalar
field potential is an evolving effective potential-- which can have
a well-defined minimum (Sanders 1989).  
Initially, the potential evolves faster
than the scalar field can respond, but at some point, the scalar
falls and oscillates about the minimum.  If the bare potential has
quadratic form these oscillations constitute CDM with, possibly,
a long Compton wavelength-- so long that these soft bosons cannot
accumulate in galaxies.  At the present epoch, 
the matter-coupling field provides a fifth force that appears below
a critical value of the scalar field gradient as in MOND.  
Cosmological CDM exists
as long wavelength excitations of the coupling-strength field.
I show that a preferred-frame generalization of PCG 
reproduces the galaxy scale phenomenology of
MOND.  I give a numerical example of cosmological evolution which
illustrates that the theory can reproduce the basic properties
of the concordance model and, in particular, is consistent with the
cosmological evidence for CDM.  I further demonstrate
that solar system phenomenology constrains the parameters of the
theory and the wavelength of the bosons, but, inevitably, 
this fifth force must be present in the outer 
solar system-- a phenomenon possibly detected
as the Pioneer effect (Anderson et al. 1998).

\section{A brief history of ideas}

The basis for the suggested scalar-tensor theories of modified dynamics
is the non-relativistic, but Lagrangian-based, modified Poisson equation of 
Bekenstein and Milgrom (1984):
$$\nabla\cdot[\mu(|\nabla\phi|/a_0)\nabla\phi]=4\pi G \rho \eqno(1)$$
where $\mu$ is the function interpolating between the Newtonian
regime ($\mu(x) = 1$) and the MOND regime ($\mu(x) = x$).
In a scalar-tensor theory, $\phi$ would refer only to the scalar
field and not to the total gravitational potential.  That is to say,
in the weak field limit, such a theory would be a two-field theory
where, in addition to $\phi$, there is the ``normal'' Newtonian field 
satisfying the usual Poisson equation.  

Phase coupling gravitation (PCG) is a covariant
scalar-tensor generalization for such modified gravity proposed by 
Bekenstein (1988a, 1988b).
Here, the scalar field is complex ($\xi = qe^{i\phi}$) 
and the field Lagrangian is standard:
$$L_s = {1\over 2}[q^2\phi_{,\alpha}\phi^{,\alpha} + q_{,\alpha}
 q^{.\alpha} + 2V(q)] \eqno(2)$$
The non-standard aspect is that only the phase couples to matter 
jointly with the gravitational tensor $g_{\mu\nu}$, i.e.,
$$L_I = L_I[exp(-2\eta\phi)g_{\mu\nu}...], \eqno(3)$$  
This leads to the scalar field equation,
$$(q^2\phi^{,\alpha})_{;\alpha} = {{8\pi \eta G T}\over{c^4}}.\eqno(4)$$
Here it is obvious that the scalar amplitude squared plays the role
of the MOND interpolating function $\mu$.  

PCG looks like a
proper field theory, with the Lagrangian possessing a self-interaction
potential $V(q)$ as well as a standard kinetic term.  
Bekenstein demonstrated that
MOND-like phenomenology in a static background 
can result if $V(q) \propto -q^6$; i.e., galaxy phenomenology
requires a ``negative sextic'' potential.  In general,
an attraction falling less rapidly than $1/r^2$ 
requires $V'(q) = dV/dq <0$ over some range of $q$ (Sanders 1988).

The theory is interesting because it can possess a viable cosmological 
limit where
$a_0$ is identified with $\eta c\dot\phi$ (Sanders 1989), and as such,
becomes a candidate for a cosmological effective theory of MOND.  
In an isotropic,
homogeneous Universe (Robertson-Walker), the equations for the cosmic
evolution of the scalar phase and amplitude (Einstein conformal frame)
are
$$\dot\phi = 3{\eta\over{q^2}}\Omega_b{a^{-3}}\tau \eqno(5)$$
and
$$\ddot q +3h\dot q = -V'(q) + q{\dot\phi}^2 \eqno(6) $$
where $\Omega_b$ is the density parameter of the baryonic
matter, $a$ is the universal scale factor in terms of the present
scale factor, time ($\tau$) is in units of ${H_0}^{-1}$
and $h$ is the Hubble parameter in units of $H_0$.  Looking at eq.\ 6
we see that the evolution of the scalar amplitude can be described
in terms of a time-dependent ``time-like'' effective potential,
$$V_t(q) = V(q) + {1\over 2}{{k(\tau)^2}\over {q^2}} \eqno(7) $$
where, from eq.\ 5, we have $$k(\tau) = 3\eta\Omega_b{a^{-3}}\tau
\eqno(8)$$
The obvious
attraction point of such a potential, $\bar q$, is a local minimum, which 
implies that the bare potential should be a monotonically increasing
function of q, as in  $V(q) = Aq^n$. 

In a Universe with inhomogeneities, the scalar phase and
amplitude are given, in the quasi-static limit by the solutions to
$$\nabla\cdot (q^2\nabla\phi)={{8\pi\eta G\rho}\over{c^2}}\eqno(9)$$
$$\nabla^2q -q\nabla\phi\cdot\nabla\phi = V'(q) - \dot\phi^2 q^2,\eqno(10)$$
where $\dot\phi$ is the cosmic time derivative of the scalar phase given
by eq.\ 5 with $q = \bar q$, its expectation value.  Thus, the solution for
q involves a second time-dependent ``space-like'' effective potential given by
$$V_s(q)=V(q)- {1\over 2}{{k(\tau)^2}\over{\bar q}^4}q^2 \eqno(11)$$
That is to say, cosmology adds a time-dependent negative mass term
to the bare potential.
Here the attraction point $\bar q'$ is at a local maximum.  If certain very
general conditions on the bare potential are satisfied, the maximum
in the space-like potential occurs at the same value of $q$ as the
minimum in the time-like potential, $\bar q' = \bar q$; i.e.,
the scalar amplitude asymptotically
approaches its cosmic value far from a mass concentration (Sanders 1989).  
A cosmological
theory can be constructed in which the total force about a mass concentration
deviates from pure Newtonian below a critical value of the total acceleration
as in MOND, but there is a return to $1/r^2$
attraction on larger scale with a maximum possible mass discrepancy
of $\delta = 2\eta^2/{\bar q}^2$.  This predicted maximum discrepancy 
is not large ($\delta \leq 10$) so the theory does not produce
extended flat rotation curves for spiral galaxies as observed; this
only seems possible in the context of Bekenstein's original $V(q)\propto
-q^6$ theory.

Apart from this phenomenological difficulty, a more fundamental problem with
PCG, or any scalar-tensor theory in which the scalar couples to 
matter as a conformal factor multiplying the Einstein
metric (eq.\ 3), is that the scalar field does not interact with photons.
This is to say, there is no enhanced gravitational lensing due
to the scalar field;  lensing by large astronomical systems should reveal no
``dark matter'' in sharp contrast to the observations (Bekenstein 
\& Sanders 1994).  This led Sanders (1997) to propose a non-conformal 
relation between the physical and Einstein metrics involving
an additional non-dynamical vector field $A^\mu$ pointing in
the positive time direction in the preferred cosmological frame,
as in the classical
stratified theories (Ni 1972).  Here the metric of the
physical geometry, $\tilde{g}_{\mu\nu}$ is related to the gravitational metric
$g_{\mu\nu}$ as
$$\tilde{g}_{\mu\nu} = 
e^{-2\eta\phi}g_{\mu\nu} - 2 sinh(2\eta\phi)A_\mu A_\nu. \eqno(12)$$
This was combined with an aquadratic Lagrangian theory (Bekenstein
\& Milgrom 1984) to produce the phenomenology of
MOND along with enhanced gravitational lensing.  

The essential problem
with this aquadratic stratified theory is the non-dynamic vector field;  
such a construct
violates the principle of General Covariance making it impossible to
define a conserved energy-momentum tensor (Jacobson \& Mattingly 2001).  
This has been remedied by
Bekenstein (2004) who has endowed the vector field with its own
dynamics and demonstrated that a theory can be constructed that is
causal, that produces MOND phenomenology about mass concentration,
that possesses a sensible cosmological limit,
and in which the relationship between the deflection of photons by
a mass concentration and the total 
weak-field force (scalar plus tensor) is the same as in General Relativity.
The scalar action of Bekenstein's tensor-vector-scalar theory, TeVeS,
is essentially that of PCG in the limit of weak coupling (small $\eta$)
where it can be shown that the kinetic term for the scalar amplitude, $q$,
can be neglected in the Lagrangian;  i.e.,
$$L_s = {1\over 2}q^2\phi_{,\alpha}\phi^{,\alpha} + V(q)\eqno(13)$$
In the weak field limit this leads to a field equation equivalent
to eq.\ 1 with $\mu(x) = q^2$ where $q$ is given
by the solution to $qx^2 = V'(q)$.  As a toy model, 
Bekenstein chooses a form for $V(q)$ which yields two discontinuous
branches for $\mu(x)$;  one for negative argument that would
be relevant to cosmology; and one for positive $x$, relevant to
quasi-stationary systems (mass inhomogeneities).

Here I am going to put back the kinetic term for $q$.  This is vital because
I will identify oscillations in $q$ with CDM, so its dynamics must be followed
fully.  The idea that long-wavelength scalar field oscillations may
comprise dark matter is also not new:  Press, Ryden \& Spergel (1990) 
proposed that such ``soft bosons'' would form after a late phase
transition (after recombination) leading to the development of 
density fluctuations that would be the seeds for structure formation.
The suggested potential provides a Compton wavelength of 30 kpc
corresponding to a supposed current scale of large scale structures (30 Mpc).
In a unified model of dark energy and dark matter,
Sahni \& Wang (1999) also propose scalar field oscillations as dark
matter and suggest that a large Compton wavelength could solve several
of the outstanding observational problems of CDM halos, such as the 
apparent absence of central density cusps and the low abundance of
dwarf galaxies in the Local Group.  They referred to such dark matter
as FCDM-- ``Frustrated Cold Dark Matter''.  This was also considered
by Hu, Barkana, \& Gruzinov (2000) who emphasize that the de Broglie wavelength
(the Compton wavelength extended by $c/v$) is the relevant clustering
scale for such `fuzzy' cold dark matter.  In the present context, the
de Broglie wavelength should be sufficiently long that the bosons do not
accumulate in galaxies at all; this could be termed SFCDM or ``seriously
frustrated cold dark matter''. 

For galaxy or solar system phenomenology, we will see that it is necessary to 
take very weak coupling ($\eta<<1$), so the scalar sector of the 
theory is very close
to that of TeVeS.  As in the stratified theory I will not consider here
the separate dynamics of the vector field;  
although, a {\it dynamical} vector
field is a necessary requirement for general covariance, 
and it must be included properly in any final theory.
It does appear, however, that the only consequence of the 
dynamical vector field in the weak field limit 
is a rescaling of the local gravitational constant, $G$, with respect to
its cosmological value (Bekenstein 2004, Carroll \& Lim 2004, Giannios
2005). 

First of all, I describe a preferred-frame 
generalization of PCG because PCG,
in a cosmological context, cannot produce the phenomenology of MOND.

\section{Bi-scalar-tensor-vector theory (BSTV)}
\subsection{The scalar field equations}
The normalized vector field, dynamical or not, takes its simplest
(time-like) form in the cosmological frame.  Therefore a 
theory involving such a field is inevitably a preferred frame 
theory.  Moreover, the unit vector, {\bf A},
may be used to form a second scalar field invariant
$$J = A^\mu A^\nu \phi_{,\mu}\phi_{,\nu} \eqno(14)$$
in addition to the usual invariant
$$I = g^{\mu\nu}\phi_{,\mu}\phi_{,\nu} \eqno(15)$$
(Sanders 1997).
Obviously $J =\dot\phi^2$ in the preferred cosmological frame, and
$$K = I+J \eqno(16)$$ becomes the square of the spatial gradient of $\phi$
in this frame ($K=\nabla\phi\cdot\nabla\phi$).  
This means that we can make use of the preferred frame to separate, 
and manipulate separately, the spatial
and time derivatives of $\phi$ at the level of the field Lagrangian.

Therefore the theory that I consider will have the general scalar field
Lagrangian,
$$L_s = {1\over 2}[q_{,\alpha}q^{,\alpha} + h(q)K - f(q)J + 2V(q)].
\eqno(17)$$
That is to say, separate functions of $q$ multiply the spatial and
temporal gradients of $\phi$ in the preferred frame.  Obviously the
fields $q$ and $\phi$ can no longer be identified with the amplitude
and phase of a complex scalar, but these fields do play the same roles
as in PCG:   $\phi$ is the matter coupling field 
 and $q$ determines the strength of that coupling.
The interaction of $\phi$ with particles is described by the
action
$$S_p = -mc \int{\Bigl[-\tilde{g}_{\mu\nu}
{{dx^\mu}\over{dp}}{{dx^\nu}\over{dp}}\Bigr]^{1\over 2} dp}\eqno(18)$$
where the physical metric $\tilde{g}_{\mu\nu}$ is given by eq.\ 12, and $p$
is a parameter along the path.
 
The scalar field dynamics comes from the action
$$S_s = {{c^4}\over{16\pi G}}\int{L_s \sqrt{-g} d^4 x}\eqno(19)$$
combined with the matter action (eq.\ 18 summed over particles).
Taking $\delta S_s = 0$ with respect to variations in $\phi$, one finds
$$\bigl(P^{\mu\nu}\phi_{,\mu}\bigr)_{;\nu} = {{8\pi G\eta}\over {c^4}}
{\tilde{T}_{\mu\nu}\bigl[g^{\mu\nu}+(1+e^{-4\eta\phi})A^\mu A^\nu\bigr]
} \eqno(20)$$ where $\tilde{T}_{\mu\nu}$ is the energy-momentum tensor in
the physical frame and  
$$P^{\mu\nu} = h(q)g^{\mu\nu} + [h(q)-f(q)]A^\mu
 A^\nu. \eqno(21)$$
For an ideal fluid this becomes 
$$\bigl(P^{\mu\nu}\phi_{,\mu}\bigr)_{;\nu}={{8\pi G\eta}\over{c^2}}
(\tilde\rho + 3\tilde{p})e^{-2\eta\phi} \eqno(22)$$
where $\tilde\rho$ and $\tilde{p}$ are the density and pressure actually
measured by a co-moving observer (see Bekenstein 2004 for a derivation of
the source term).  We notice that, unlike usual scalar-tensor
theory, photons and other relativistic particles act as a source for $\phi$.
We also notice that in the preferred frame $P^{\mu\nu}$ takes 
a particularly simple
form with the space-space ($i,j$) and time-time (0,0) components 
given by
$$P^{i,j} = h(q)g^{i,j},\,\,\,\,\,\,\,\, P^{0,0} = -f(q) \eqno(23)$$

Setting $\delta S_s/\delta q = 0$  yields the field equation for $q$:
\setcounter{equation}{23}
\begin{eqnarray}
\lefteqn{-(g^{\mu\nu}q_{,\mu})_{;\nu} +  
{1\over 2}h'(q)\phi_{,\alpha}\phi^{,\alpha}} \nonumber
 \\ & & -{1\over 2} [h'(q) -f'(q)]A^\mu A^\nu \phi_{,\mu}\phi_{,\nu} + 
V'(q) = 0 
\end{eqnarray}
which, in the preferred frame, simplifies to
$$(g^{\mu\nu}q_{,\mu})_{;\nu} - {1\over 2} h'(q) \nabla\phi\cdot\nabla\phi
 + {1\over 2} f'(q) {\dot\phi}^2 - V'(q) = 0\eqno(25)$$   

The complete theory includes the Einstein-Hilbert action for the
tensor field and, in principle, an action for the vector field,
but here the separate dynamics of the vector is not included.
Hereafter, all operations are carried out in the preferred frame because 
this is where the equations assume their least complicated form.
 
General empirical considerations constrain the form of the free functions
of the theory: $h(q)$, $f(q)$, and $V(q)$.  First of all, assuming
homogeneity and that $e^{-2\eta\phi}\approx 1$ (justified below), the cosmic
time derivative of $\phi$ in the matter-dominated epoch is 
found by integrating eq. 22:
$$\dot\phi = {{-3\eta\Omega_b\tau {a^{-3}}}\over{f(q)}}.\eqno(26)$$
Thus, in a cosmological setting ($\nabla q = 0$, $\nabla\phi = 0$), 
the evolution of $q$ (eq.\ 25) is determined by a
time-like effective potential (as in PCG) 
$$V_t(q) = V(q) + {1\over 2}{{{k(\tau)}^2}\over {f(q)}} \eqno(27)$$
with $k(\tau)$ again given by eq.\ 8. 
I wish to identify
dark matter with $q$ oscillations in this effective potential.  The fluid
represented by such oscillations has an equation of state
$p=w\rho$, and, for a power law bare potential, it is straightforward to
demonstrate that 
$w = (n-2)/(n+2)$  (Turner 1983).  
Identification of 
the scalar field oscillations with CDM ($w=0$) requires $n=2$; therefore,
I take $$V(q) = {1\over 2}Aq^2 + B.\eqno(28)$$  Although the theory
contains an evolving component of dark energy (quintessence), it will be 
seen below that
inclusion of the bare cosmological term $B$ is necessary to match the
concordance model.

The cosmological expectation value of $q$ (i.e., $\bar q$)
is given by the solution to
$${{dV_t}\over {dq}} = A\bar{q} - {1\over 2}
{{{{k(\tau)}^2}f'(\bar q)}\over{{f(\bar q)}^2}} = 0. \eqno(29)$$
To be an attractor, this extremum in the potential must also be a
minimum which requires that ${V_t}''(\bar q) >0$.
This condition, combined with eq.\ 29, provides a general condition
on $f(q)$; i.e., $$ {d\,{ln(f'/f^2)}\over {d\,{ln(q)}}} < 1 \eqno(30)$$
evaluated at $\bar q$.

One may question the use of such a time-dependent effective potential
with a well defined minimum because $V_t$ and $q$ are both 
functions of time.
But the two timescales are generally very different;
the expectation value $\bar q$ varies on a 
cosmic timescale while the period of
oscillation about $\bar q$ is constant and fixed by the
bare potential ($\propto 1/\sqrt{A}$).  As long as the oscillation period
is much shorter than the age of the Universe (not true in the early
Universe) the concept of a time-like effective potential is
meaningful.

In a Universe with inhomogeneities, but where $\dot q$ can be neglected,
we may view eq.\ 25 as containing a space-like potential
(as in PCG) which determines the spatial dependence of $q$:
$$V_s(\tau) = {1\over 2}Aq^2 + B - {1\over 2}
{{{k(\tau)}^2}\over{{f(\bar q)}}^2}f(q).\eqno(31)$$
The extremum at $\bar q'$, given by the solution to 
${{dV_s}/{dq}} = 0$, occurs
at the same value of $q$ as the extremum in $V_t$; i.e., $\bar q = \bar q'$.
But the extremum in this case must be a local maximum in $V_s(q)$; i.e.
${V_s}''(\bar q) < 0$, which implies a second condition on $f(q)$; i.e.,
$${{d\,{ln(f')}}\over{d\,{ln(q)}}} > 1 \eqno(32)$$
evaluated at $\bar q$.
If conditions 30 and 32 are met, then the cosmological boundary condition
on $q$ is satisfied; i.e., far from a mass concentration $q$ approaches
its cosmological value.

One choice for the free functions which satisfies these
conditions would be
$h(q) = q^2$ and $f(q) = q^6$.  This is appealing because it 
would provide a cosmological
realization of Bekenstein's negative sextic effective potential in 
static PCG and hence MOND phenomenology on an extragalactic scale.  In fact,
such a model leads to a scalar force rising as $1/r$ well into a
galactic mass distribution; this fifth force,
in addition to the Newtonian gravitational force, produces not flat
but declining rotation curves in the outer parts of spiral galaxies.
Moreover, this theory violates precise constraints on $1/r^2$ attraction
in the solar system.  These considerations require a slight modification
of the free functions:
$$h(q) = {\epsilon^2}(1-e^{-q^2/\epsilon^2}) \eqno(33)$$
and $$f(q) = {{q^6}\over{{\epsilon^2}(q^4+\eta^4)}}.\eqno(34)$$
Here, in addition to the coupling strength parameter $\eta$, a second
parameter $\epsilon$ ($\approx 100 \eta$) has been introduced.

It may appear that the presence of three free functions,
$V(q)$, $h(q)$, and $f(q)$, characterized by three free parameters,
$A$, $\eta$, and $\epsilon$, introduces a great deal of arbitrariness into 
the theory. In fact, given the structure of the theory (eq.\ 17), 
the free functions must have forms similar
to eqs.\ 28, 33, and 34.  The potential must be quadratic to assure
that $w=0$ for the dark matter component; in the extragalactic limit,
$q<\eta$, it must be the case that $h(q)\rightarrow q^2$ and
$f(q)\rightarrow q^6$ to yield MOND phenomenology in the outskirts of
galaxies (as in PCG); and in the inner solar system, it necessary that
$h(q)\rightarrow \epsilon^2 >> \eta^2$ to assure negligible scalar field
effects and precise inverse square
attraction.  This is a theory strongly constrained by phenomenology.

\subsection{BSTV as MOND}

The quasi-static BSTV field equation 
for $\phi$ in the presence of mass inhomogeneities (eq.\ 22 with 23) is: 
$$\nabla\cdot[h(q)\nabla\phi] = {{8\pi G\eta \rho}\over{c^2}}; \eqno(35)$$
when $q<\epsilon$, this becomes equivalent to that of static
PCG, eq.\ 9 ($h(q)\approx q^2$).

The scalar force in the low-velocity limit is given by
$$f_s = \eta c^2 \nabla\phi. \eqno(36)$$
Then we find, by identification of eq.\ 35 with eq.\ 1, that the MOND
interpolating function $\mu$ is equivalent to 
$q^2/2\eta^2$.  In the limit
of weak coupling, where $q_{,\alpha} q^{,\alpha}$ may be neglected in the 
Lagrangian as in eq.\ 13, and where ${\bar q}<q<\eta$ 
(the outskirts of galaxies) we may neglect the first and final
terms in eq.\ 25.  With eqs.\ 33 and 34, this yields 
$$\mu = {{q^2}\over{2\eta^2}} = {\epsilon\over{\sqrt{12}}}
\sqrt{{\nabla\phi\cdot\nabla\phi}
\over{\dot\phi^2}}. \eqno(37)$$  
That is to say, in this limit the relation between $q$ and $\nabla\phi$
becomes algebraic (as in TeVeS).
Eq.\ 37 then implies that the MOND acceleration
parameter should be identified as $$\alpha = {{a_0}\over{cH_0}}
= {{\sqrt{12}\,\eta|\dot\phi|}\over{\epsilon}}
\eqno(38)$$ where $\dot\phi$ is in units of
the inverse Hubble time.  
In this theory, as in PCG, there is a clear cosmological 
origin of $a_0$ which is identified with the cosmic time derivative of
the matter-coupling scalar field, $\dot \phi$.  The acceleration parameter
evolves with cosmic time, and I take $\alpha_0$ as its current value.

Far beyond a bounded mass distribution of total mass M, the scalar force
(eq.\ 36) approaches $$f_s = {\delta GM\over {r^2}}\eqno(39)$$ where 
 $$\delta = {{2\eta^2}\over{\bar q^2}}.\eqno(40)$$  
That is to say, the total force, scalar
plus Newtonian, becomes $1/r^2$ with a ratio of the total
to Newtonian force of $1+\delta$.  Because of the cosmic evolution
of $\bar q$ it is evident that $\delta$ also evolves; 
I take $\bar q_0$ and $\delta_0$ to be the present values.

\begin{figure}
\resizebox{\hsize}{!}{\includegraphics{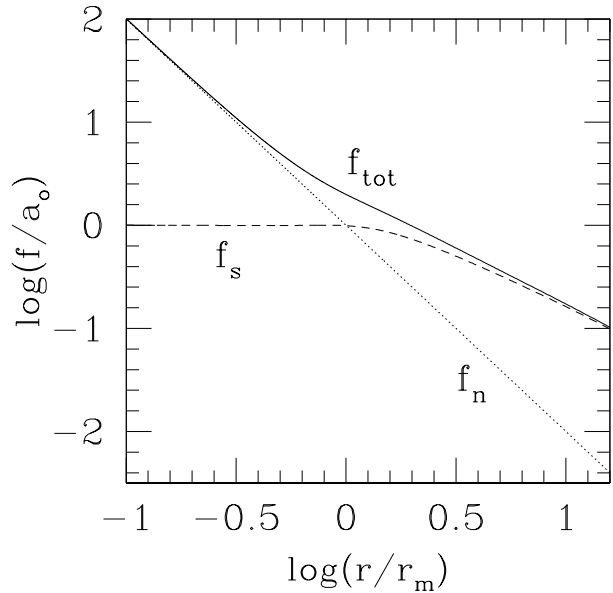}}
\caption[]{The log of the Newtonian force, $f_n$, scalar force, $f_s$,
and total force, $f_t$, in units of $a_0$ plotted against the log of
the radial distance from a point mass in units of the MOND radius
$\sqrt{GM/a_0}$.  This results from numerical solution of the BSTV
field equations with $\delta_0 = 50$ and $\alpha_0 = 0.125$.
 }
\end{figure}

The full solution for $q$ and $\phi$ about a mass concentration involves 
solving eq.\ 35 simultaneously with eq.\ 25 which, in this limit,
may be written as
$$\nabla^2 q - h'(q)\nabla\phi\cdot\nabla\phi = V'_s(q)\eqno(41)$$
where the space-like effective potential is 
$$V_s(q) = {1\over 2} (Aq^2 - f(q){\dot\phi}^2) + B \eqno(42)$$
with $h(q)$ and $f(q)$ given by eqs. 33 and 34.

The basic parameters of the theory are the scalar mass-squared $A$,
the coupling constant $\eta$, and an additional parameter $\epsilon$ 
which has no effect on galactic fields so long as $\epsilon>100\eta>q$.
The procedure followed here is to take values of the MOND acceleration
parameter and maximum mass discrepancy, $\alpha_0$ and $\delta_0$, that
are consistent with galaxy phenomenology in the context of MOND
(i.e.,  $\delta_0>25$ and $\alpha_0<0.2$) and then to determine
the parameters of the theory that are required by these
phenomenological constraints.  In fact, the values of $A$ and
$\eta$ are not specified by these considerations but are related
through the condition that ${V_s}'(\bar q_0) = 0$. Given eq.\ 42
this becomes
$$A= {1\over 2}f'(\bar q_0){{{\dot{\phi_0}}^2}\over{\bar q_0}} =
{1\over 2}{{\dot{\phi_0}}^2} {{f(\bar q_0)}\over{\bar{q_0}}^2}
{{d\,ln(f)}\over{d\,ln(q)}} \approx \Bigl({{\alpha_0}\over {{\delta_0}
{\eta}}}\Bigr)^2.\eqno(43)$$ where, in the approximation, I have 
taken $f(q)\approx q^6/(\epsilon^2\eta^4)$ 
in the limit where $q<\eta$ (the extragalactic limit) 
and made use of eq.\ 38 and 40.
Therefore, it is necessary to assume a value of either $A$ or $\eta$,
and here I take $\eta=3\times 10^{-8}$ to be consistent with weak
coupling (more physical constraints follow below).  
With $\delta_0 = 50$, and $a_0= 0.125{cH_0}$, then $A=6.9\times 10^9$ 
${H_0}^2$ (this would correspond to a mass of about $10^{-28}$ ev).

With these parameters, the solution of eqs.\ 35 and 41 in the presence 
of a point mass yields the scalar force $f_s/a_0$ shown in
Fig.\ 1 as a function of radius, in units of the MOND radius $r_m=
\sqrt{GM/a_0}$.  Also shown are the Newtonian force and
the total force.  We notice that when $f_{tot}/a_0<1$ the scalar force
exceeds the Newtonian force, and 
the total force falls as $1/r$ rather than $1/r^2$

The resulting rotation curves of spherical galaxies having the
density distribution of a Plummer sphere (Binney \& Tremaine 1987) 
and total masses of $10^{11}$ M$_\odot$ and $ 3\times 10^{10}$ M$_\odot$
are shown in Fig.\ 2.  In both cases the core radius of the sphere
is 2 kpc.  It is evident
that such a theory can reproduce those essential properties of MOND
which so nicely embody the overall characteristics of 
galaxy rotation curves:  no Keplerian decline beyond the visible disk;
an extended flat rotation curve with the characteristic rotational velocity 
proportional to
the one-fourth power of the galaxy mass (the Tully-Fisher law);
a larger mass discrepancy in systems with lower surface density;
and a different overall form of rotation curves in high- and
low-surface density systems.  A difference with MOND in its 
original form is an eventual Keplerian decline in rotation velocity and,
consequently, the presence of an upper limit to the mass discrepancy of
$\delta+1$, but this can be large.

\begin{figure}
\resizebox{\hsize}{!}{\includegraphics{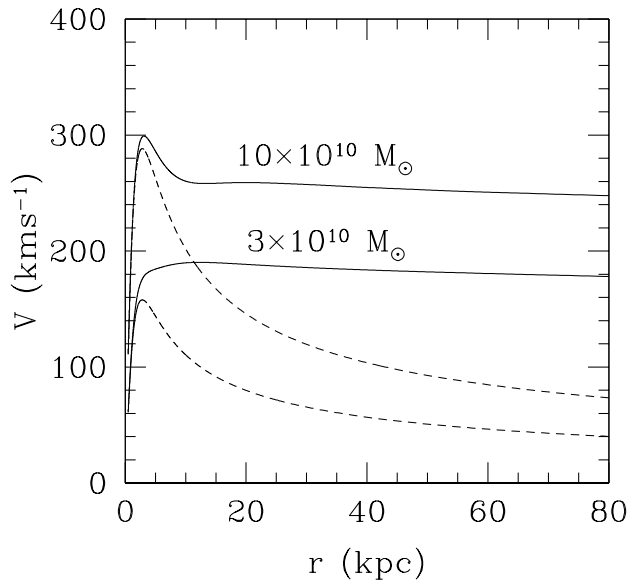}}
\caption[]{The rotation curves of spherical galaxies (Plummer spheres)
having total masses of $10^{11}$ and $3\times 10^{10}$ M$_\odot$ and
a core radius of 2 kpc in both cases.
The solid lines show the rotation curves including the scalar force and
the dashed lines show the Newtonian rotation curves.  As in Fig.\ 1
this results from solution of the BSTV field equation for $\delta_0=50$
and $\alpha_0=0.125$
 }
\end{figure}

Clearly this is a cosmologically effective theory of MOND;
that is, the form of rotation curves is determined by the effective
potential eq.\ 42 which depends upon the cosmology via $\dot\phi$.  
But $\dot\phi$ is related to $\Omega_b$ and the parameters of the
theory as given in eq.\ 26; with eqs.\ 38 and 40 and taking $q<\eta$
this is
$$\alpha_0 = {{3\sqrt{3}}\over 4}\Omega_b\epsilon{\delta_0}^3 .\eqno(44)$$
Both from considerations of primordial nucleosynthesis and the observed
CMB anisotropies it is known that $\Omega_b \approx 0.04-0.05$.
Therefore, if $\alpha_0<0.2$ and $\delta_0 > 25$ as is required for 
galaxy phenomenology, it must be the case that $\epsilon$ 
is small.  In the above example, ($\delta_0=50$,
$\alpha_0=0.125$) we find that $\epsilon\approx 2\times 10^{-5}$ if
$\Omega_b\approx 0.04$.  

I now demonstrate that the theory delivering this MOND phenomenology can also
be consistent with concordance cosmology and solar system phenomenology.

\subsection{BSTV as CDM}

The full time-like effective potential (eq.\ 27) is given by
$$V_t(q) = {1\over 2}Aq^2+B+{1\over 2}{k(\tau)}^2{\epsilon^2
{(q^4+\eta^4)}\over{q^6}}.\eqno(45)$$
The cosmological expectation value of $q$, given by the condition 29, is
$$\bar q^4 = {{k(\tau)^2\epsilon^2}\over {2A}}\Bigl[1+\Bigl\{1+12 {{
A\eta^4}\over{ (k(\tau)\epsilon)^2}}\Bigr\}^{1\over 2}\Bigr].\eqno(46)$$
Given the form of $k(\tau)$ is evident that $\bar q$ decreases as
the Universe ages.

The actual value of $q$ will oscillate about this expectation value
and is given by the solution to eq.\ 25; in the cosmological (homogeneous)
limit this becomes  
$$\ddot q + 3h\dot q = -{V_t}'(q)\eqno(47)$$
where ${V_t}'$ is determined from eq.\ 45.
The Einstein field equation for the gravitational metric remains the same 
but the two scalar fields
contribute to the energy-momentum tensor.  Thus the Friedmann equation for
the expansion of the Universe becomes
$$h^2 = \Omega_r a^{-4} + \Omega_b e^{-2\eta\phi}a^{-3} +
{1\over 6}[{\dot q}^2 + f(q){\dot\phi}^2 + 2V(q)]\eqno(48)$$ 
where $\Omega_r$ is the density parameter of radiation, and I have 
assumed that $\Omega_{total} = 1$.

The response time
of the scalar field to the effective potential
is comparable to the oscillation period at the
bottom of the time-like potential well; i.e., for small oscillations,
$$\tau_r = {{2\pi}\over{\sqrt{V_t''(\bar q)}}}\eqno(49)$$
where $$V_t''(\bar q) = A\Bigl[1+{{d\,ln(f^2(\bar q)/f'(\bar q))}\over {d\,
ln(\bar q)}}\Bigr]\eqno(50)$$
Then, from eq.\ 43 it follows that,
$$\tau_r \approx 2.2 {{\delta_0\eta}\over {\alpha_0}}\eqno(51)$$
For the parameters yielding the MOND phenomenology, described above,
($\eta= 3\times 10^{-8}$, $A= 6.9\times 10^9 {H_0}^{2}$) this is
$\approx 2.6\times 10^{-5}$ ${H_0}^{-1}$ 
(about $3.2\times10^5$ years if $H_0=75$).  While the Universe
is younger than $\tau_r$, the scalar field cannot respond to the
evolving effective potential.  The scalar field $q$ will remain at its
initial value until $\tau\approx \tau_r$, will then seek the 
effective potential minimum and oscillate about that value.  
The oscillations would thus constitute dark matter, in the form of
soft bosons, with a Compton wavelength of about $c\tau_r \approx 110$ kpc. 

\begin{figure}
\resizebox{\hsize}{!}{\includegraphics{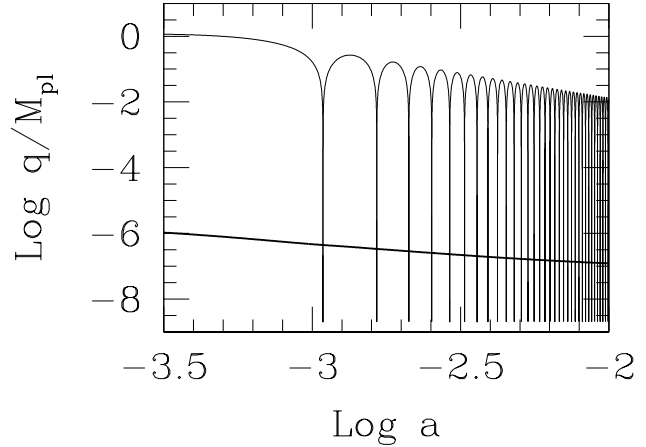}}
\caption[]{A log-log plot of the scalar field, $q$, as a function of 
scale factor, $a$, normalized to the present scale factor, for the 
cosmological model described in the text.  The heavy solid line shows
the cosmic evolution of the expectation value, $\bar q$ (the bottom of
the potential well).  For $a<10^{-3}$ the response time of the
scalar field to the effective potential is longer than the age of the
Universe.  
 }
\end{figure}

The evolution of a model Universe with these properties has been followed
by numerically integrating eqs.\ 26, 47, and 48.  
Here I have set $\Omega_b = 0.04$ which requires $\epsilon = 2\times
10^{-5}$ to be consistent with the supposed MOND phenomenology
($\alpha_0 = 0.125$, $\delta_0 = 50$).
In integrating eqs.\ 47 and 48 it is assumed that the scalar field 
oscillations remain
coherent down to the present epoch.  The results are 
shown in Figs. 3-5.  In Fig. 3
we see the early evolution of the scalar field $q$ as a function of scale
factor $a$. This illustrates the point made above:  for $a\geq  
10^{-3}$, corresponding to an age of roughly $3\times 10^{-5}$ ${H_0}^{-1}$,
the response time of the scalar field to the effective potential becomes
less than the age of the Universe;  $q$ feels the potential and seeks its
minimum.  But it overshoots the equilibrium point and runs against the
``centrifugal'' barrier ${k(\tau)}^2/f(q)$.  The oscillations persist through
the present epoch; their contribution 
to the present density of the Universe depends
upon the amplitude which depends upon the initial value of the
the scalar field $q$.  Here it was necessary to set this at 1.2 $M_{pl}$
(Planck mass)
to achieve $\Omega_{CDM}\approx 0.3$.  The time-dependent density
parameters of radiation ($\Omega_r$), baryons ($\Omega_b$), the oscillating
component of the scalar field ($\Omega_{dm}$) and the smooth component of
the scalar field ($\Omega_\lambda$) are shown in Fig.\ 4. 

In this theory there is an evolving component of the vacuum energy
(quintessence) equal roughly to the bare quadratic potential 
$V \approx A{\bar q}^2$.  With
$A$ given by eq.\ 43 and making use of eq.\ 40, we find that
$V \approx {{\alpha_0}^2}/{\delta_0}^3$; i.e.,
the evolving component cannot comprise a large
fraction of the {\it present} total energy density 
and be consistent with MOND.   
Therefore, I have set the bare cosmological term $B= 2.1$ to force 
the present composition of
the Universe to be that of the concordance model ($\Omega_\lambda 
= 0.7$).

Over the course of the evolution of the model Universe, from 
$a=10^{-5}$ to $a=2$, the quantity $2\eta\phi$ changes from its
initial assumed value of zero to $-1.4\times 10^{-3}$.  Therefore
the approximation $exp(-2\eta\phi) \approx 1$ in determination of
$\dot\phi$ (eq.\ 26) appears to be justified.  This also implies that
the effective value of the constant of gravity $G$ has not varied by
more than a 0.5\% over the history of the Universe, so there is no
contradiction with the standard Big Bang nucleosynthesis.

\begin{figure}
\resizebox{\hsize}{!}{\includegraphics{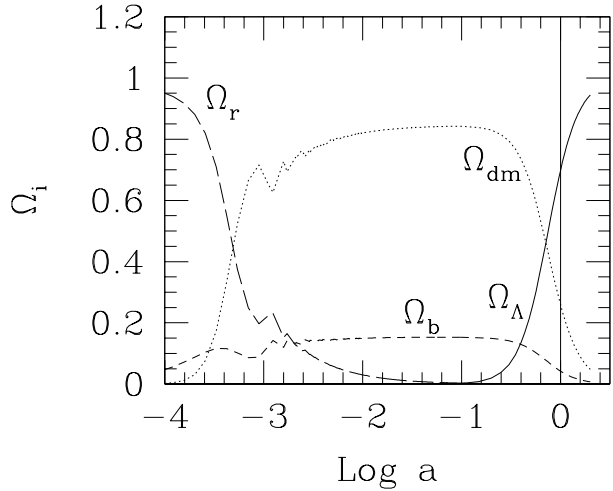}}
\caption[]{The running density parameter of the various constituents of
the Universe as a function of log $a$.  Shown are $\Omega_r$ (radiation),
$\Omega_b$ (baryons), $\Omega_{dm}$ (dark matter in the form of scalar field
oscillations) and $\Omega_{\Lambda}$ (vacuum energy density). The solid
vertical line marks the present epoch where the composition 
matches that of concordance cosmology.  
 }
\end{figure}

However, because of the cosmic evolution of $\bar q$ and $\dot\phi$, 
those aspects of
galaxy kinematics related to MOND, the acceleration parameter, $\alpha$, 
and the maximum mass discrepancy, $\delta$, will evolve with cosmic
time.  This is shown in Fig.\ 5 where we see these quantities as a function
of scale factor.  The change in the MOND critical 
acceleration, $\alpha$, would be noticeable as an evolution of 
asymptotic rotation velocity and consequently of the
Tully-Fisher relation for spiral galaxies.  For example, at a redshift
of $z=1$, the acceleration parameter is roughly half its present value
which implies that the asymptotic circular velocity corresponding
to an object of a given mass would be only 85\% of its present value.

\begin{figure}
\resizebox{\hsize}{!}{\includegraphics{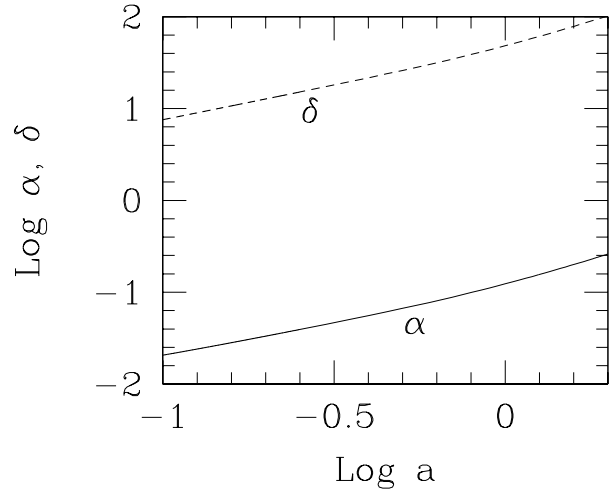}}
\caption[]{A log-log plot of the cosmic evolution of the MOND acceleration
parameter, $\alpha = a_o/cH_0$ (solid line), and the maximum mass
discrepancy, $\delta$ (dashed line).  The MOND phenomenology would be
clearly evident from a redshift of 2 through the present epoch; although
the asymptotically constant value of a galaxy rotation velocity would
increase with cosmic time.
 }
\end{figure}

The Compton wavelength of the 
bosonic dark matter, in units of the Hubble length, is given by
$c\tau_r$ (eq.\ 51).
Taking $h_0=0.75$ this may be written 
$$\lambda_c = 8.8 \times 10^6  {{\delta_0 \eta}\over{\alpha_0}}\,\,\,
  {\rm kpc} \eqno(52)$$  

The observations of the CMB anisotropies place an upper limit on
$\lambda_c$ because the scalar dark matter must form well 
before recombination ($\tau_r<3\times 10^{-5}$).
If the soft bosons are to cluster on at least the
scale of the third peak in the CMB power spectrum, then
$\lambda_c < 40$ kpc.  

The condition that the bosons not collect in galaxies
places a lower limit on $\lambda_c$. 
The minimum clustering scale is the de Broglie wavelength 
given by $l_c = \lambda_c (c/v)$ where v is the velocity dispersion
of a bound object ($v\approx 300$ km/s for a massive galaxy).  
If, for example, $l_c>50$ kpc it must be the case that $\lambda_c>
.05$ kpc.  

These two conditions take together restrict the range
of $\eta$;  from eq.\ 52 we find 
$$ 7\times 10^{-12} \leq \eta({0.125/\alpha_0}) ({\delta_0/100}) 
\leq 6\times 10^{-9}
\eqno(53)$$

If $\eta$ lies within these limits,
then the bosons could collect in clusters of 
galaxies ($l_c\leq 1$ Mpc) while avoiding galaxies.  
This is a possible solution to 
the problem of the remaining mass discrepancy in clusters in the context
of MOND (Sanders 2003).  

Below we see that  
solar system phenomenology places an even more stringent upper limit on 
$\eta$ and hence, via inequality 53, a lower limit on the maximum mass 
discrepancy $\delta_0$.

\section{Solar System Constraints}

The solution of the BSTV field equations about a galaxy-scale
($10^{11}$ M$_\odot$),
albeit spherical, mass distribution,
as shown in Fig. 1, implies that $q \approx 2.2 \eta$ at the 
approximate position of the sun ($r=8$ kpc), corresponding to a  
local discrepancy of $\delta \approx 0.4$.
In other words, in the outer solar system there is
a scalar field acceleration that is comparable to the 
Newtonian acceleration.

In the inner solar system, the avoidance of locally detectable
preferred frame effects limits the contribution of a scalar field 
to the total force (Sanders 1997, in preparation 2005); specifically, 
the acceleration due to the scalar field gradient
must be substantially smaller than the Newtonian
acceleration, i.e., $f_s<10^{-4}f_n$.  Moreover, within the 
orbit of Neptune,
planetary motion restricts the total weak field force, Newtonian plus 
scalar, to be quite precisely inverse-square, such that
any spatial variation of Kepler's constant, $GM_\odot$, is less than 
several parts in
about $10^5$ (Anderson et al. 1995).  Therefore, the substantial
scalar field acceleration
present in the Galaxy at the radius where a test particle would begin
to feel the acceleration of the sun (about 1 pc) must become
significantly smaller than the Newtonian force 
within the orbit of Neptune.  This provides, from the following
argument, an upper limit on the scalar field coupling $\eta$.

In the limit where $q<\epsilon$ the quasi-static field equation for 
q about a point mass (eq.\ 41) may be written 
$$\nabla^2 q = q\bigl({{2\eta GM}\over{c^2 q^2 r^2}}\bigr)^2 - 
    {1\over 2} f'(q){\dot\phi}^2\eqno(54)$$
Defining $y=q/\bar q_0$ and $x = r/r_m$ where $r_m = 
\sqrt{GM/a_0} = \sqrt{r_h r_s/\alpha_0}$
($r_s$ is the Schwarzschild radius and $r_H$ is the Hubble radius) 
and making use of eqs.\ 38 and 40, this may be rewritten as
$${{2\eta^2}\over{\alpha_0\delta_0}}{{r_H\over r_s}{\nabla_x}^2 y} = 
{1\over 2}{{\delta_0}\over{y^3 x^4}} - {1\over{6\delta_0}} y.\eqno(55)$$ 
Here for the final term I have assumed that $q>\eta$ ($f'(q) \approx
2q/\epsilon^2$) which is generally the case within the Galaxy.

Eq.\ 55 may be used to estimate
the radius at which a massive object begins to affect the variation
of $q$.  Setting $\nabla^2 y \approx 1/x^2$ (i.e., $\Delta y = 1$) we
find a critical radius given by
$$r_c \approx {1\over 2} {{\delta_0}\over{\eta}} r_s.\eqno(56)$$
This means that the anomalous force about the sun would
increase as $1/r^2$ into $r_c$ within which the mass of the sun would
affect the radial dependence of $q$, leading to a non-inverse square force. 
The requirement that this deviation from $1/r^2$ be unnoticeable in the
inner solar system places a lower limit on $r_c$ (for example 1000 a.u.)
and therefore a revised upper limit on $\eta$; i.e.,
$$\eta<10^{-11} \delta_0 \bigl({M\over {M_\odot}}\bigr)^{0.5} 
\bigl({{1000\,\, a.u.}\over {r_c}}\bigr)
\eqno(57)$$.

\begin{figure}
\resizebox{\hsize}{!}{\includegraphics{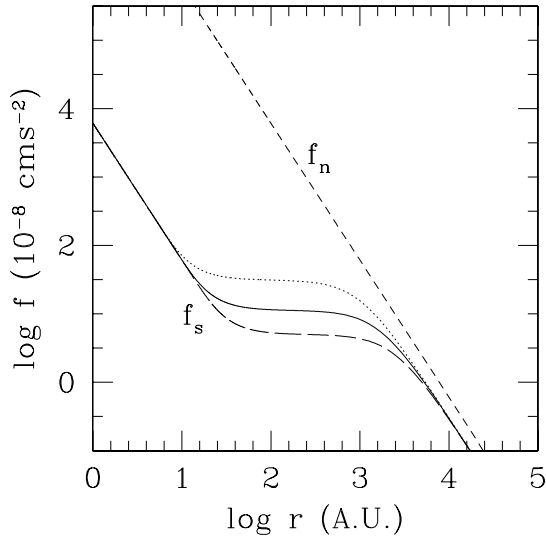}}
\caption[]{A log-log plot of the Newtonian and scalar accelerations
(in units of $10^{-8}$ cm/s$^2$)
about a 1 M$_\odot$ point mass as a function of distance.  The dashed
line is the Newtonian acceleration and the dotted, solid, and long-dashed
curves are the scalar accelerations corresponding respectively to 
$\eta = 5\times 10^{-12}$, $3\times 10^{-12}$, and $2\times 10^{-12}$.
In all cases $\epsilon=141\eta$}
\end{figure}

Fig.\ 6 illustrates the numerical solution to the BSTV field
equations (eqs.\ 35 and 41) 
about a 1 $M_\odot$ object with $\bar q_0= {2\eta}$ ($\delta_0 = 0.4$)
at infinite radius
(the star is embedded in the Galaxy).  I have taken three different values
of $\eta$ (2, 3, and 5 $\times 10^{-12}$).  In all cases $\epsilon = 
{141}\eta$.  The dashed line on the 
log-log plot is the Newtonian force and the long dashed, dotted and solid
curves show, respectively, the anomalous scalar
force for these three different assumed values of $\eta$.  
We see that in all cases the anomalous force increases as $1/r^2$ into about 
1000 A.U. and then becomes constant at a value between 4 and 100 $\times 
10^{-8}$ cm/s$^2$ depending upon the assumed value of $\eta$.
The scalar force remains constant until $q\rightarrow \epsilon$ 
at which point the
scalar force is given by $f_s = {({2\eta^2}/\epsilon^2) f_n}$; i.e.,
at smaller radii the force is again $1/r^2$ but a factor of 
$2\eta^2/\epsilon^2$ below the Newtonian force.  This, in fact, is an
additional motivation for the parameter $\epsilon$ as given in eq.\ 33.

It would appear that a constant force in the outer solar system is 
unavoidable in the context of this theory. 
This is interesting in view of the anomalous
acceleration detected in the outer solar system by the 
Pioneer spacecrafts-- 
the ``Pioneer anomaly''-- which is on the order of 
$10^{-7}$ cm/s$^2$ (Anderson et al. 1998). If the predicted
anomalous force
is to be less than the Pioneer anomaly, then $\eta<3\times 10^{-12}$.
This upper limit, combined with the lower limit given by the 
condition that the soft bosons should not accumulate in galaxies,
inequality\ 53, places a limit upon the maximum discrepancy 
in the present universe; i.e.,  $\delta_0\geq 2000$ if $\eta \leq 
3\times 10^{-12}$.  Thus the predicted galaxy phenomenology approaches that of
pure MOND with a very large maximum discrepancy.

In this limiting case, corresponding to $\lambda_c \approx 50\,\,pc$, the
response time of the scalar field would be about 160 years; i.e.,
scalar field oscillations would appear when the Universe approximately
200 years old or at $a = 5\times 10^{-6}$, well before decoupling.
This would seem to guarantee that the soft bosons could cluster down
to the scale of the third or fourth peak in the CMB angular power spectrum,
although detailed calculations would be required to confirm
consistency with the observed power spectrum.  This limit would also
assure that the bosons could penetrate to the cores of rich clusters
of galaxies.

As in all scalar-tensor theories there is a predicted time
variation of the baryonic mass unit or, equivalently, of 
the gravitational constant.  In this case, it arises because
the source term for the gravitational metric, $g_{\mu\nu}$,
contains the expression $exp(-2\eta\phi)$.  Therefore,
the cosmological variation of G would
be given by $$\dot G/G = -\eta\dot\phi = 2\epsilon\alpha_o H_0\eqno(58)$$
Given that the MOND acceleration parameter $\alpha_o \leq 0.2$ 
and that $\epsilon \approx 200 \eta$, then with the above limit
on $\eta$ this would
imply that $\dot G/G \approx 10^{-11}$ $H_0$,  well below the current
limit imposed by 
lunar laser ranging; i.e., $\dot G/G<0.1$ $H_0$ (Williams et al. 2004).

\section{Conclusions}

The near numerical coincidence between the MOND acceleration parameter and
$cH_0$ suggests that the proper theory of MOND should be an effective
theory-- that MOND phenomenology results only
in a cosmological background.  If so, then the goal is not to
derive a relativistic theory which predicts, on the one hand,
MOND phenomenology about mass concentrations, and on the other hand, 
leads to a viable cosmology in a homogeneous universe.
Rather, the correct theory may well be one in which 
MOND reflects the influence of cosmology on local
particle dynamics and arises only in a cosmological setting.

It goes without saying that this theory is not General
Relativity, because in the context of General Relativity
local particle dynamics is immune to the influence of cosmology.
This may be demonstrated via the Birkoff theorem (Will \& Nordtvedt
1972), but ultimately comes down to the strict application of
the Equivalence Principle (the Strong Equivalence Principle) in
GR which permits no environmental influence on local particle dynamics
(apart from tides).  But, as was first fully appreciated by Dicke (1962), 
this does not apply to scalar-tensor theory where the cosmology-encoded
scalar field, determined by the universal mass distribution and its
time evolution, pervades every corner of the Universe.  This suggests
that a cosmological basis for MOND may be provided by scalar-tensor theory.

However, as was also appreciated by Dicke, 
if scalar-tensor theory is to satisfy the precise experimental constraints
on the universality of free fall (the Weak Equivalence Principle), then
it must couple to matter jointly with the Einstein metric.  
The simplest such coupling, via a conformal
transformation of the Einstein metric, leads to the serious 
problem of gravitational
lensing in the context of MOND-- no enhanced deflection of photons
due to the scalar field.  This problem may be solved by postulating a 
non-conformal relation between the Einstein and physical metrics
involving a normalized dynamical vector field (as in TeVeS), but such a field
may also produce unobserved effects of the preferred 
cosmological frame on local particle dynamics.

The PCG field equation, or its weak coupling limit, provides a mechanism 
by which unwanted local
dynamical effects of the scalar or vector fields may be suppressed
in the limit of high field gradients (e.g., in the solar system) where
the proper theory of gravity is very nearly GR.  In this context
the outskirts of
galaxies would be the transition region between a preferred frame 
tensor-vector-scalar cosmology and a GR-dominated local dynamics that is
protected from cosmological influence.  This 
transition would be observable as an acceleration dependent deviation
from Newtonian dynamics, or MOND.  

I have shown 
here that a preferred frame generalization of PCG, i.e. a bi-scalar-tensor 
vector theory (BSTV), can be an effective theory of MOND
in the sense that it provides a cosmological realization of Bekenstein's
negative sextic potential in the context of static PCG.
Moreover, it is a cosmologically evolving 
effective potential, and the MOND phenomenology can only result in the context
of a FRW cosmology.  The theory is characterized by three parameters:
in addition to the mass-squared of the bare potential, $A$, and the coupling
strength parameter, $\eta$, it was necessary to add a third parameter,
$\epsilon$ which
permits a sensible value of $\Omega_b$ in the presence of MOND 
phenomenology (eq.\ 44), provides a return to $1/r^2$ attraction in the
inner solar system (Fig.\ 6), and tames the rapid evolution of $G$
(eq.\ 58).

The evolving potential also provides a mechanism which,
with the quadratic bare potential, inevitably produces cosmological cold dark 
matter in the form of scalar field oscillations.  This unavoidable
appearance of cosmological dark matter is a primary motivation 
for the theory.  Such dark matter appears to be required by a variety
of considerations ranging from CMB anisotropies to the observed
relative dimming and re-brightening of SNIa.  

However, if the dark matter is not to
accumulate in galaxies, then it must be seriously frustrated; i.e.,
the Compton wavelength, extended by $c/v_{rot}$ 
should be larger than the scale of galaxies.
I have demonstrated that this implies a lower limit to the strength
of the scalar field coupling to baryons (eq.\ 53)-- a limit which also
depends upon the maximum ratio of the scalar-to-Newtonian forces (the
maximum discrepancy).
There is also an upper limit on the coupling parameter required by
the suppression of a non-inverse square 
anomalous force in the outer solar system-- that is to say, any non-inverse 
square force must be smaller than that implied by the Pioneer anomaly or by
planetary motion.  However, in view of the Pioneer anomaly, it is of 
interest that the theory requires, at some level, a constant 
anomalous acceleration in the outer solar system.  

I emphasize that the present theory is not an alternative
to Bekenstein's TeVeS;  it is an example of a TeVeS theory, but one which
explicitly involves two scalar fields; in Bekenstein's theory, the
second field is implicit (or not explicitly dynamical).  
However, this bi-scalar-tensor-vector theory is
incomplete.  The complete covariant theory requires consideration of the
full dynamics of the vector field, and, then, following Bekenstein,
an examination of the properties of wave propagation. There remains
the danger of acausal propagation or instability of the background
solution.  I leave this for later consideration because, as is 
evident, the procedure followed in constructing this theory, and that of 
Bekenstein, is different from what is usually done in relativity and
cosmology.  Here, a theory is built up from the bottom by adding bits
and pieces as required by phenomenology.  This may seem
unfamiliar or cumbersome because the usual methodology in this 
field is more deductive;  here the approach to the final theory is 
incremental.
   
It may be that the basis of MOND lies
in another direction entirely--  as a modified non-local particle action
(Milgrom 1994).  It may also be that the observational case for CDM has 
been overstated,
as argued by McGaugh (2004b).  But the weight of present 
observational evidence implies that the final theory of MOND should reproduce,
or at least simulate, the effects of CDM upon cosmic expansion and upon
CMB anisotropies.  

I am grateful to Jacob Bekenstein and Moti Milgrom for helpful 
criticisms and comments which had a major positive impact on this work.
I also thank Stacy McGaugh for helpful comments on the manuscript.

\end{document}